\documentclass[%
reprint,
nofootinbib,
 amsmath,amssymb,
 aps,
]{revtex4-2}

\usepackage{graphicx}
\usepackage{dcolumn}
\usepackage{bm}
\usepackage{xfrac}
\usepackage{hyperref}
\usepackage[mathlines]{lineno}

\usepackage[caption=false]{subfig}
\usepackage{makecell}

\usepackage{booktabs} 
\newcommand{\di}{\mathrm{d}}

\newlength{\sublen}
\begin{document}

\preprint{APS/123-QED}

\title{A Dynamical-Time Framework for the Dynamics of Charged Particles}

\author{Zui Oporto}
\email[Corresponding author: ]{zoporto@fiumsa.edu.bo}
\affiliation{Instituto de Investigaciones Físicas, and Planetario Max Schreier, Universidad Mayor de San Andrés, Campus Universitario, C. 27 s/n Cota-Cota, 0000 La Paz, Bolivia.}

\author{Gonzalo Marcelo Ram\'{\i}rez-\'Avila}
\email{gonzalo-marcelo.ramirezavila@unamur.be}
\affiliation{Namur Institute for Complex Systems (naXys), Universit\'e de Namur, Rue de Bruxelles 61, B-5000 Namur, Belgium.}
\affiliation{Instituto de Investigaciones Físicas, and Planetario Max Schreier, Universidad Mayor de San Andrés, Campus Universitario, C. 27 s/n Cota-Cota, 0000 La Paz, Bolivia.}




\date{\today}

\begin{abstract}

We present a dynamical framework for modeling the motion of point-like charged particles, with or without mass, in general external electromagnetic fields. A key feature of this formulation is the treatment of time coordinate as a dynamical variable. The framework applies to the relativistic regime while consistently admitting a nonrelativistic limit. We also introduce a representation of particle trajectories in velocity space, which provides clear insight into the nature and asymptotic behavior of the dynamics. As an application, we compare the motion of massive and massless particles in a constant electromagnetic field and find that, for identical field configurations, their asymptotic behavior is independent of both mass and initial conditions. Finally, we explore the computational advantages of the dynamical-time formulation over the conventional uniform-time approach in two case studies: a uniform electromagnetic field, and an elliptically polarized wave propagating along a uniform magnetic field. In both scenarios, the proposed scheme exhibits improvements in accuracy and computational efficiency.
\end{abstract}
\maketitle

\section{Introduction}

The possible existence in nature of fundamental particles that are both massless and electrically charged has not been ruled out from a theoretical perspective, despite the absence of experimental evidence. Nevertheless, effective massless charged particles emerge in various physical contexts. Prominent condensed matter analogs include graphene \cite{geimGrapheneExploringCarbon2007,castronetoElectronicPropertiesGraphene2009} and Weyl semimetals \cite{hosurRecentDevelopmentsTransport2013}, where electronic excitations obey a linear energy-momentum dispersion relation. Similarly, in plasma physics, models employing a fluid of massless electrons have been developed \cite{buchnerSpacePlasmaSimulation2003,buchnerSpaceAstrophysicalPlasma2023}. Furthermore, a formal analogy has been established between the dynamics of massless charged particles and that of a non-Hermitian two-level quantum system~\cite{botetDualityNonHermitianTwostate2019}.

In general relativity, the propagation of massless particles (such as photons or, hypothetically, gravitons) in the weak-field limit can be described within a framework in many respects analogous to electrodynamics. In this approach, the energy of the particles plays the role of a gravitational charge, coupled to external gravitoelectromagnetic fields~\cite{ruggieroTaleAnalogiesReview2023}.

According to relativistic theory, spinless massless particles travel at the speed of light along null geodesics and, if electrically charged, experience geodesic deviation in external electromagnetic fields, while always maintaining a constant speed equal to $c$ \cite{moralesBehaviourChargedSpinning2018}.

A substantial body of literature addresses the classical dynamics of electrically charged particles, with or without spin and irrespective of their mass; a comprehensive review, however, lies beyond the scope of this work. A partial survey is provided in Ref.~\cite{moralesBehaviourChargedSpinning2018}. For complementary approaches not covered there, see Refs.~\cite{Plyushchay:1988ws,Plyushchay:1988wx,Nesterenko:1991av}, as well as the extensive bibliography in Ref.~\cite{arreaga-garcia2014equations}. The dynamics of massive particles is well understood; for a complete analysis we refer to Ref.~\cite{Deriglazov2017}. Of particular relevance to the present study are the Lagrangian formulations in Refs.~\cite{brinkLocalSupersymmetrySpinning1976,brinkLagrangianFormulationClassical1977,balachandranClassicalDescriptionParticle1977}, in which the kinetic term is decoupled from the mass, naturally encompassing the massless case.

In nonlinear dynamics, the behavior of charged particles in electromagnetic fields represents a fruitful area of research \cite{akhiezerDynamicsHighenergyCharged1995,ramDynamicsChargedParticles2010}. For instance, massive charged particles in a dipolar magnetic field exhibit dynamical behavior ranging from periodic and quasiperiodic to chaotic and hyperchaotic in both the low-velocity \cite{dealcantarabonfimChaoticHyperchaoticMotion2000} and relativistic regimes \cite{shebalinStormerRegionsAxisymmetric2004,xiePeriodQuasiperiodChaos2020}. Extending such studies to massless particles could therefore provide valuable insights into ultrarelativistic dynamics. However, a primary difficulty in pursuing this extension is that, formally, the Lorentz factor diverges as the particle mass approaches zero, thereby challenging the applicability of standard approaches~\cite{petriFullyImplicitNumerical2017}.

A further difficulty in formulating relativistic theories lies in the inherent presence of constraint equations in phase space. While some of these constraints simply remove redundant degrees of freedom, others additionally serve as generators of gauge symmetries \cite{henneauxQuantizationGaugeSystems1992}. In either case, their presence introduces non-physical degrees of freedom that must be properly handled.

The numerical integration of constrained dynamical systems may begin with a phase-space reduction, achieved by gauge-fixing and solving the constraint equations to eliminate non-dynamical degrees of freedom. Nevertheless, this reduction is not always feasible in practice. An alternative approach is to impose a gauge condition to partially remove some degrees of freedom and proceed with integration without explicitly solving all constraints. This latter approach demands that: (i) the initial conditions must satisfy the constraints and (ii) the numerical evolution preserves them within a desired tolerance. Particular attention is required for systems sensitive to perturbations that drive the evolution off the constraint submanifold of the full phase space, as such deviations can be amplified, leading to unstable and physically meaningless results even in the short term. Special care is therefore essential, as no physical meaning can be ascribed to a simulation that fails to preserve the constraints of the system over time.

This work aims to develop a general framework that unifies the treatment of both massive and massless charged particle dynamics in external electromagnetic fields. Although related approaches have been extensively explored, the scheme proposed here is distinctive in that it treats the dynamics as a constrained system and incorporates both the energy and the temporal coordinate among the dynamical variables. In doing so, it avoids numerical instabilities caused by the Lorentz factor, which becomes extremely large for massive particles near the speed of light \cite{petriFullyImplicitNumerical2017} and is not even defined for the massless case.


In Section~\ref{sec:Dynamics of charged particles}, we begin by briefly reviewing the Lagrangian and Hamiltonian formalisms for charged particles, covering both massive and massless cases. We then introduce in Section~\ref{sec:A framework with dynamical time} a constrained dynamical framework in which the temporal coordinate is treated as an additional dynamical degree of freedom and provide a brief discussion of its low-velocity limit. In the same section, we reformulate the initial-value problem by introducing the concept of the ``velocity sphere''. Numerical applications of the proposed scheme are presented in Section~\ref{sec:Estudios-de-caso} for both uniform and time-dependent electromagnetic fields. Finally, Section~\ref{sec:Conclusions} summarizes our main results and discusses their implications.

We adopt the mostly negative Minkowski space-time metric, $\eta_{\mu\nu} = \mathrm{diag}(1,-1,-1,-1)$. The coordinates in this space are denoted by $x=\{x^{\mu},\mu=0,\ldots,3\}$, where Greek indices label four-dimensional components. When necessary, we separate the coordinates according to the convention $t = x^0$ and $\mathbf{x} = \{x^a, a=1,\dots,3\}$, with Latin indices denoting spatial components. Throughout the text, we employ the Einstein summation convention for repeated indices. Unless otherwise stated, we work in natural units $c = 1$, where $c$ is the speed of light in vacuum.

\section{Dynamics of charged particles
\label{sec:Dynamics of charged particles}}

In this section, we review the classical dynamics of charged particles, largely following Refs.~\cite{moralesBehaviourChargedSpinning2018,moralesQuantumChargedSpinning2019}, whose conventions we adopt.

The dynamics of a particle with mass $m$ and charge $q$, coupled to an external electromagnetic field $A_{\mu}$, can be described by the action
\begin{equation}
S[x]=-\int_{\mathcal{C}}\di\lambda\left(m\sqrt{\dot{x}^{2}}+qA_{\mu}(x)\dot{x}^{\mu}\right).\label{eq:action-ng}
\end{equation}
Here, $\dot{x}^{2}=\eta_{\mu\nu}\dot{x}^{\mu}\dot{x}^{\nu}$, where the dot denotes differentiation with respect to the evolution parameter $\lambda$. This action is invariant under global Lorentz transformations and, being homogeneous in the velocities, it is also manifestly gauge invariant under general reparametrizations of $\lambda$.
 
In the limit $m\rightarrow 0$, the kinetic term vanishes, resulting in a Lagrangian with linear velocity terms. This possibility was explored for a particular planar dynamics in Ref. \cite{dunneTopologicalChernSimonsQuantum1990}, and more generally in Ref. \cite{saavedraDegenerateDynamicalSystems2001}, where the degenerate nature of the dynamical system was elucidated.

Based on general symmetry principles, \citet{brinkLocalSupersymmetrySpinning1976} proposed an action describing the dynamics of a charged spinning particle, with or without mass. For the present discussion, we omit the degrees of freedom associated with spin and write the simplified action as (see \cite{moralesBehaviourChargedSpinning2018} for further details)
\begin{equation}
\tilde{S}[x,e]=-\int_{\mathcal{C}}\di\lambda\left(\frac{\dot{x}^2}{2e}+\frac{em^2}{2}+qA_{\mu}(x)\dot{x}^{\mu}\right),\label{eq:action-Brink}
\end{equation}
where $e(\lambda)$ is a Lorentz-scalar, non-physical degree of freedom with units of inverse mass. This field ensures the invariance of the action under reparametrizations $\lambda\rightarrow\lambda'(\lambda)$, under which it transforms as a worldline scalar density, $e'(\lambda')\mathrm{d}\lambda'=e(\lambda)\mathrm{d}\lambda$. With $\varepsilon(\lambda)$ an infinitesimal parameter, reparametrization invariance is realized as a gauge symmetry of the action, with transformation rules
 \begin{equation}
   \delta_\varepsilon x^\mu = \varepsilon \dot{x}^\mu\,,\quad \delta_\varepsilon e = \dot{\varepsilon} e+\varepsilon \dot{e}\,,\quad \delta_\varepsilon A_\mu = \varepsilon \dot{A}_\mu\,.
\end{equation}

In the action~\eqref{eq:action-Brink}, the mass is decoupled from the kinetic term and appears as a free parameter multiplying the length of the particle’s worldline. As a consequence, the massless sector is realized in a straightforward manner. Moreover, one may even perform the substitution $m^2 \rightarrow -m^2$ to describe tachyonic particles.\footnote{The action~(\ref{eq:action-Brink}) can be interpreted as a theory of (super)gravity in $0+1$ dimensions coupled to a multicomponent scalar field $x^\mu(\lambda)$~\cite{brinkLocalSupersymmetrySpinning1976,brinkLagrangianFormulationClassical1977}. In this interpretation, $m^2$ plays the role of a cosmological constant, which may take negative values under the substitution $m^2 \to -m^2$.}

Applying functional variations to the action \eqref{eq:action-Brink} with respect to $x^\mu$ and $e$, we obtain, respectively,
\begin{subequations}\label{eq:eom_covariant_0}
\begin{align}
\frac{\di}{\di\lambda}\left(e^{-1}\dot{x}^{\mu}\right) & =qF^{\mu}\!_{\nu}\dot{x}^{\nu},\label{eq:ma-LF}\\
\dot{x}{{}^2}-m^{2}e^{2} & =0\,,\label{eq:mass-shell}
\end{align}
\end{subequations}
where $F_{\mu\nu}=\partial_\mu A_\nu - \partial_\nu A_\mu$ is the electromagnetic field strength tensor. The first equation describes the dynamics of the charged particle, independent of its mass; the second establishes an algebraic relation between the velocity components and the mass, signaling the presence of constraints. Note that the cases $m\neq0$ and $m=0$ define two distinct sectors of the theory that must be treated separately: In the massive case, the constraint can be solved to eliminate the auxiliary variable $e$, while for massless particles, this is not possible, leaving $e$ as a pure gauge degree of freedom.

The Hamiltonian formulation of this theory, in its most general form describing a charged spinning particle, was developed in Ref.~\cite{moralesQuantumChargedSpinning2019}. In the present case, where spin is neglected, the first-order action takes the form
\begin{equation}
\tilde{S}[e,x,p]=\int \di\lambda\left(p_{\mu}\dot{x}^{\mu}-eH(x,p)\right),\label{eq:action-star-H}
\end{equation}
where $p_{\mu}\equiv \delta \tilde{S}/\delta \dot{x}^\mu$ are the canonical momenta computed from the action~\eqref{eq:action-Brink}, and the Hamiltonian is given by
\begin{equation}
H(x,p)=\frac{1}{2}\left(m^{2}-\eta^{\mu\nu}(p_{\mu}+qA_{\mu})(p_{\nu}+qA_{\nu})\right)\,.\label{eq:hamiltonian}
\end{equation}

Variation of the action \eqref{eq:action-star-H} with respect to $e$ yields the Hamiltonian constraint equation
\begin{equation}
H(x,p)\approx 0\,,
\end{equation}
where, following the Dirac-Bergmann formalism for singular systems \citep{henneauxQuantizationGaugeSystems1992}, we use the weak equality, $\approx$, which prescribes that the constraints must be treated as strong equalities only after all Poisson brackets have been evaluated.

For the canonically conjugate pair, satisfying $\{x^{\mu},p_{\nu}\}=\delta^\mu_\nu$, the Hamilton equations yield the following dynamical system:
\begin{align}
\dot{x}^{\mu} & = \{x^{\mu},e H\}=-e\eta^{\mu\nu}\left(p_{\nu}+qA_{\nu}\right),\label{eq:dot-X}\\
\dot{p}_{\mu} & = \{p_{\mu}, e H\}=qe\eta^{\nu\sigma}\partial_{\mu}A_{\nu}\left(p_{\sigma}+qA_{\sigma}\right).\label{eq:dot-P}
\end{align}

When $A_{\mu}$ is independent of certain coordinates, say $x^{\hat{\mu}}$, Eq.~(\ref{eq:dot-P}) implies that the corresponding canonical momentum is an integral of motion. We reserve the symbol $\mathcal{P}_{\hat{\mu}}$ for these conserved momenta.

This completes the description of phase-space dynamics.

The Hamiltonian formalism provides a natural framework for studying the evolution of the system. However, since the equations of motion depend on the electromagnetic potentials, the components \(p_{\mu}\) are not themselves physical observables. Nevertheless, the dynamics become well defined once a gauge is fixed. In the next section, we present a reformulation of the system that is manifestly $U(1)$ gauge-invariant.


\section{A framework with dynamical time\label{sec:A framework with dynamical time}}

Let us introduce the \emph{kinetic momentum} of the particle as $P^{\mu}\equiv \dot{x}^{\mu}/e$, with $e^{-1}$ assuming the role of inertia. The relation with canonical momenta is given by
\begin{equation}
    p_{\mu}=-\eta_{\mu\nu}P^{\nu}-qA_{\mu}\,.
\end{equation}
In terms of the kinetic momenta variables Eqs.~\eqref{eq:eom_covariant_0} rewrite as
\begin{subequations}\label{eq:eom_covariant}
\begin{align}
\dot{x}^{\mu} & = eP^{\mu}\,,\label{eq:dot-xx}\\
\dot{P}^{\mu} & = \frac{q}{c}F^{\mu}\!_{\nu}eP^{\nu}\,,\label{eq:dot-pp}\\
\chi & \equiv P^{2}-m^{2}=0\,.\label{eq:mass-shell-01}
\end{align}
\end{subequations}
 In these equations, $e$ remains a gauge degree of freedom. The right-hand side of Eq.~(\ref{eq:dot-pp}) corresponds to the Lorentz force, which makes the dynamical system manifestly $U(1)$ gauge-invariant. Moreover, the \emph{mass-shell constraint}, denoted by $\chi$, is now independent of $e$.

A quantitative analysis of the dynamics requires gauge-fixing the variable $e$. To this end,  we first introduce the notation $x^\mu\equiv(t,\mathbf{x})$ and $P^\mu\equiv(\mathcal{E},\mathbf{P})$. From this, the zeroth component of Eq.~(\ref{eq:dot-xx}) reads  $\dot{t}=e(\lambda)\mathcal{E}(\lambda)$.

We define the \emph{uniform-time gauge} by the condition $e(\lambda)=1/\mathcal{E}(\lambda)$, which is equivalent to fixing $\dot{t}=1$, hence the name of this gauge.

We define the \emph{dynamical-time gauge} by the condition $e(\lambda)=1/\mathcal{E}(0)$, which is equivalent to fixing $\dot{t}=\mathcal{E}(\lambda)/\mathcal{E}(0)$. So, for this gauge, the time coordinate evolves proportionally to the energy of the particle.

For massive particles, the uniform-time gauge allows the energy to be expressed as $\mathcal{E}=m \gamma(\dot{\mathbf{x}})$, where $\gamma(\dot{\mathbf{x}})=1/\sqrt{1-\dot{\mathbf{x}}^2}$ is the Lorentz factor. In this gauge, $\mathcal{E}$ is no longer an independent variable but is determined by the velocity of the particle. This procedure is equivalent to performing a reduction in the dimensionality of the dynamical system.

In the dynamical-time gauge, $\mathcal{E}(0)$ denotes the energy of the particle evaluated at $\lambda=0$. For massive particles, we choose the initial condition $\mathcal{E}(0)=m$, which yields $\dot{t}=\mathcal{E}(\lambda)/m$; consequently, the evolution parameter $\lambda$ coincides with the particle’s proper time.

We should note that the elimination of $\mathcal{E}$ from the set of dynamical variables, or the use of the particle’s proper time as the evolution parameter, cannot be achieved when the mass of the particle is zero.  In what follows, we will adopt the dynamical-time gauge, $e(\lambda)=1/\mathcal{E}(0)$, irrespective of the mass of the particle.

After splitting spatial and time coordinates, the set of Eqs.~\eqref{eq:eom_covariant} becomes \cite{ProccIWOSP}
\begin{subequations}\label{eq:DTF}
\begin{align}
\dot{t}             & = \frac{\mathcal{E}}{\mathcal{E}_{0}}\,, \label{eq:DTF-01} \\
\dot{\mathbf{x}}    & = \frac{\mathbf{P}}{\mathcal{E}_{0}}\,, \label{eq:DTF-02} \\
\dot{\mathcal{E}}   & = \frac{q}{\mathcal{E}_{0}} \mathbf{E}(t,\mathbf{x})\!\cdot\!\mathbf{P}\,, \label{eq:DTF-03} \\
\dot{\mathbf{P}}    & = \frac{q}{\mathcal{E}_{0}} \left( \mathbf{E}(t,\mathbf{x})\mathcal{E} - \mathbf{B}(t,\mathbf{x})\!\times\!\mathbf{P} \right)\,, \label{eq:DTF-04} \\
\chi                & \equiv \mathcal{E}^{2} - \mathbf{P}^{2} - m^{2} = 0\,. \label{eq:DTF-05}
\end{align}
\end{subequations}
where $\mathcal{E}_0\equiv\mathcal{E}(0)$ denotes the initial energy. It is noteworthy that in this formulation the equations for the variables $\mathcal{E}$ and $\mathbf{P}$ are linear, with coefficients that generally depend on the coordinates $(t,\mathbf{x})$. We refer to this system as the Dynamical-Time Framework (DTF) to distinguish it from the Uniform-Time Framework (UTF), which is based on the uniform-time gauge. 

\subsection{The Low-Velocity Limit}
\label{sec:sub:The Low-Velocity Limit}

The physical content of the dynamical system remains unchanged under any choice of gauge. It should be emphasized that adopting the dynamical-time gauge is unrelated to any physically observable phenomena, such as time dilation in special and general relativity. To clarify this point, in this section, we determine the appropriate low-velocity limit. For the analysis that follows, it is convenient to restore the constant $c$ in all equations.

The low-velocity limit necessarily applies to massive particles. For these, the mass-shell constraint implies $e = \sqrt{\dot{x}^{2}}/m c$. Substituting this result into action (\ref{eq:action-Brink}) gives back the original action~(\ref{eq:action-ng}). We then separate the temporal component from the spatial ones and perform an expansion in powers of $\dot{\mathbf{x}}/c$, yielding:
\begin{align}
S[\mathbf{x},t] & =\!\int\! \di\lambda \! \left( \frac{1}{2} m \frac{\dot{\mathbf{x}}^{2}}{\dot{t}} - q\!\left(\!\dot{t} A_{0} - \frac{\dot{\mathbf{x}}}{c}\!\cdot\!\mathbf{A}\!\right) + \mathcal{O}\left( \dot{\mathbf{x}}/c \right)^{2} \right)\,. \label{eq:action-NR}
\end{align}
This action is invariant, up to a total derivative, under the following global Galilean transformations \cite{LeBellac:1973unm}
\begin{align}
    (t',\mathbf{x}') &= (t,\mathbf{x}-\mathbf{u}t)\,,\\
    (A_0',\mathbf{A}') &= \left(A_0-\mathbf{u}\!\cdot\!\mathbf{A}/c,\mathbf{A}\right)\,.
\end{align}
Also, the action is manifestly invariant under general reparameterizations $\lambda \rightarrow \lambda'(\lambda)$. We conclude that treating the temporal coordinate as a dynamical variable is not exclusive to relativistic theories.\footnote{For an insightful discussion of reparametrization invariance and its relation with time as a dynamical variable, see \citet{kieferQuantumGravity2004}.}

Applying to the action~\eqref{eq:action-NR} a procedure similar to that which led to Eqs.~\eqref{eq:DTF}, we obtain:
\begin{subequations}
\begin{align}
\dot{t}             & =f\,,\\
\dot{\mathbf{x}}    & =f\frac{\mathbf{P}}{m}\,,\\
\dot{\mathcal{E}}   & =f\frac{q}{m}\mathbf{E}(t,\mathbf{x})\!\cdot\!\mathbf{P}\,,\\
\dot{\mathbf{P}}    & =f\frac{q}{mc}\left(\mathbf{E}(t,\mathbf{x})mc-\mathbf{B}(t,\mathbf{x})\!\times\!\mathbf{P}\right)\,,\\
\chi                &\equiv\mathcal{E}-\frac{\mathbf{P}^{2}}{2m}=0\,.
\end{align}
\end{subequations}
Here, we introduce $f=f(\lambda)$ as a monotonic, yet otherwise completely arbitrary, function of $\lambda$. We note that $\chi$ corresponds to the low-velocity limit of the relativistic mass-shell constraint. In this limit, the dynamical subsystem defined by $(t,\mathbf{x},\mathbf{P})$ decouples from $\mathcal{E}$, allowing the constraint to be treated as a strong equality, $\mathcal{E}=\mathbf{P}^{2}/2m$, thus rendering the equation for $\dot{\mathcal{E}}$ redundant.

\subsection{The velocity sphere
\label{sec:sub:The velocity sphere}}

The quantity $\di\mathbf{x}/\di\lambda$ is not invariant under a general reparametrization, therefore cannot be regarded as a physical observable. Instead, the physical velocity of the particle is determined by the invariant expression $\mathbf{v}=\mathbf{P}/\mathcal{E}$, or equivalently $\mathbf{v}=\dot{\mathbf{x}}/\dot{t}$. Hence, the identification $\dot{\mathbf{x}}=\mathbf{v}$ acquires physical meaning only in the gauge $\dot{t}(\lambda)=1$.

The mass-shell constraint imposes distinct velocity-space geometries depending on the particle type. For massless particles, it requires $\mathbf{v}^{2} = 1$, defining a unit 2-sphere; physically admissible orbits are confined to this surface, so not all components of $\mathbf{v}$ can vanish simultaneously (a massless particle can never be at rest). For massive particles, introducing $\mu \equiv m / \mathcal{E}$ yields $\mathbf{v}^{2} + \mu^{2} = 1$, which defines a unit 3-sphere;\footnote{We have $\mu \in (-1,1)$ if negative energies are allowed. For strictly positive energies, this reduces to a hemisphere or, equivalently, a solid three-dimensional ball.} a physical state on this sphere is described by $u = (\mathbf{v}, \mu)$. Here, unlike the massless case, the spatial velocity can vanish when $u = (\boldsymbol{0}, 1)$,  representing a particle at rest. Finally, the formal substitution $m^{2} \rightarrow -m^{2}$ leads to $\mathbf{v}^{2} - \mu^{2} = 1$, which describes a 3-hyperboloid that corresponds to the velocity space of superluminal particles.

In a given reference frame, for a massive particle whose values of $\mathcal{E}$ and $\mathbf{P}$ are increasing arbitrarily, orbits on the velocity 3-sphere will asymptotically approach the equator, $u=(\mathbf{v},0)$. Thus, a highly energetic massive particle will asymptotically behave like a massless one. Operationally, this is equivalent to taking $m/\mathcal{E} \rightarrow 0$ (rather than $m\rightarrow0$); however, it is important to note that this limit is discontinuous, since a true massless particle has no rest frame, while a massive particle always has one. This discontinuity has a topological origin: the velocity space for massless particles is a 2-sphere, while for massive particles it is a 3-sphere, and the former cannot be continuously deformed into the latter.

The velocity sphere provides a natural parametrization of initial conditions consistent with the mass-shell constraint. Let us define $\mu_0 = m / \mathcal{E}_0$ and introduce spherical coordinates $(\varphi_0, \psi_0)\in[0,\pi]\times[0,2\pi]$ on this space. Then one has
\begin{equation}
\mathbf{v}_0 = \sqrt{1 - \mu_{0}^{2}}
\left(\sin\varphi_{0}\cos\psi_{0},\sin\varphi_{0}\sin\psi_{0}, \cos\varphi_{0}\right)\,,
\label{eq:velocidad-inicial}
\end{equation}
which automatically satisfies $\mathbf{v}_0^{2} + \mu_{0}^{2} = 1$.
Therefore, a physically admissible initial state is fully determined by the set $(\mathcal{E}_0,\varphi_0,\psi_0)$, together with the initial position $\mathbf{x}_0\equiv\mathbf{x}(0)$, regardless of whether the particle is massive or massless.


\section{Charged-particle dynamics in the dynamical-time framework \label{sec:Estudios-de-caso}}

We now examine the dynamics of charged particles within the Dynamical-Time Framework (DTF) and compare the results with those obtained using the Uniform-Time Framework (UTF). The electromagnetic configurations considered in this study are: (i) a constant field, and (ii) a monochromatic electromagnetic wave, elliptically polarized, superposed with a constant magnetic field along the propagation direction.

\subsection{Constant electromagnetic field
\label{sec:sub:Campo-electromagnetico-constante}}

Previous works on the classical dynamics of charged particles in constant external fields have shown that, despite its simplicity, this configuration already exhibits essential features, such as the $\mathbf{E}\times\mathbf{B}$ drift and the competition between $\mathbf{E}$ and $\mathbf{B}$ fields governing the long term behavior of charged particles \cite{eli_1974,takeuchiSatoshi2002,friedmanRelativisticAccelerationCharged2005}.

Consider a constant electromagnetic field defined by $\mathbf{E} = E_{0}\hat{\mathbf{e}}$, $\mathbf{B} = B_{0}\,\hat{\mathbf{z}}$, with $\hat{\mathbf{e}}=\sin\alpha\,\hat{\mathbf{y}} + \cos\alpha\,\hat{\mathbf{z}}$ and $\alpha$ the relative angle between $\mathbf{E}$ and $\mathbf{B}$. Introducing the normalized fields $\boldsymbol{\omega}_{E} = q \mathbf{E} / \mathcal{E}_{0}$ and $\boldsymbol{\omega}_{B} = q \mathbf{B} / \mathcal{E}_{0}$, along with the dimensionless variables $\mathbf{Y} = \mathbf{P} / \mathcal{E}_{0}$ and $Z = \mathcal{E} / \mathcal{E}_{0}$, the dynamical system, Eqs.~\eqref{eq:DTF}, reduces to:
\begin{subequations}\label{eq:DTF-adim}
\begin{align}
\dot{t}       & = Z\,, \\
\dot{\mathbf{x}}     & = \mathbf{Y}\,, \\
\dot{Z}       & = (\omega_{E} \sin\alpha) Y_{2} + (\omega_{E} \cos\alpha) Y_{3}\,, \label{eq:dot-Z} \\
\dot{Y}_{1}   & = \omega_{B} Y_{2}\,, \label{eq:dotY1} \\
\dot{Y}_{2}   & = (\omega_{E} \sin\alpha) Z - \omega_{B} Y_{1}\,, \label{eq:dot-Y2} \\
\dot{Y}_{3}   & = (\omega_{E} \cos\alpha) Z\,, \label{eq:dot-Y3} \\
\chi          & \equiv Z^{2} - \mathbf{Y}^{2} - \mu^{2}_0 = 0\,. \label{eq:constraint}
\end{align}
\end{subequations}
The last five equations define a closed dynamical subsystem in the variables  $(Z,\mathbf{Y})$. This structure suggests a solution strategy: first solve this reduced system, and then reconstruct the full trajectory $(t(\lambda), \mathbf{x}(\lambda))$, using the integrals of motion
\begin{subequations}\label{eq:integrals_Q}
\begin{align}
Q_{0} & = Z - (\omega_{E} \sin\alpha) x_{2} - (\omega_{E} \cos\alpha) x_{3}\,, \\
Q_{1} & = Y_{1} - \omega_{B} x_{2}\,, \\
Q_{2} & = Y_{2} + \omega_{B} x_{1} - (\omega_{E} \sin\alpha) t\,, \\
Q_{3} & = Y_{3} - (\omega_{E} \cos\alpha) t\,.
\end{align}
\end{subequations}
However, to preserve the generality of our approach for less symmetric fields, we choose to solve the full dynamical system. The constraint equation and integrals of motion are thus used to assess the numerical accuracy of our simulations.

\subsubsection{Fixed Points on the Velocity Sphere}

In terms of the velocity $\mathbf{v}=\mathbf{P}/\mathcal{E}$,  Eqs.~\eqref{eq:dot-Z}--\eqref{eq:dot-Y3} can be combined into a nonlinear equation of the form  $\dot{\mathbf{v}}=\mathbf{f}(\mathbf{v})$, where
\begin{equation}
\mathbf{f}(\mathbf{v}) \equiv -\boldsymbol{\omega}_B\times\mathbf{v}-\left(\boldsymbol{\omega}_E\times\mathbf{v}\right)\times\mathbf{v}\,.    
\end{equation}
For the field configuration with $B_{0}\neq0$ and $E_{0}\neq0$, the condition $\mathbf{f}(\mathbf{v})=\boldsymbol{0}$ yields two fixed points in velocity space, denoted by $\mathbf{v}^\pm \equiv \mathbf{v}(\pm\infty)$, whose components are given by
\begin{subequations}\label{eq:vel-space-FixPo}
\begin{align}
v_{1}^{\pm} & =\frac{\mathcal{E}_{\mathrm{EM}}-\sqrt{\mathcal{E}_{\mathrm{EM}}^{2}-\mathcal{P}_{\mathrm{EM}}^{2}}}{\mathcal{P}_{\mathrm{EM}}}\;, \\
v_{3}^{\pm} & =\pm\frac{1}{B_{0}}\sqrt{-\mathcal{F}+\sqrt{\mathcal{E}_{\mathrm{EM}}^{2}-\mathcal{P}_{\mathrm{EM}}^{2}}}\;,\\
v_{2}^{\pm} & = \pm\sqrt{1 - (v_{1}^{\pm})^{2} - (v_{3}^{\pm})^{2}}\;,
\end{align}
\end{subequations}
where $\mathcal{E}_{\mathrm{EM}}$, $\boldsymbol{\mathcal{P}}_{\mathrm{EM}}$ and $\mathcal{F}$ represent the electromagnetic energy density, Poynting vector, and field invariant, respectively:
\[
\mathcal{E}_{\mathrm{EM}}=\frac{\mathbf{E}^{2}+\mathbf{B}^{2}}{2}\,,\quad\boldsymbol{\mathcal{P}}_{\mathrm{EM}}=\mathbf{E}\times\mathbf{B}\,,\quad\mathcal{F}=\frac{\mathbf{E}^{2}-\mathbf{B}^{2}}{2}\,.
\]
The other cases of interest are summarized in Table~\ref{tab:fixed_points}.

\begin{table}[t] 
  \centering
  \begin{tabular}{@{}l p{5.3cm}@{}} 
    \toprule
    \textbf{Field condition \hspace{0.5cm}} & \textbf{Fixed points $\mathbf{v}^{\pm}$} \\
    \midrule
    $B_0 = 0$ & $(0, \pm \sin\alpha, \pm \cos\alpha)$ \\
    \addlinespace[0.5em]
    $\mathbf{B} \parallel \mathbf{E}$ & $(0, 0, \pm 1)$ \\
    \addlinespace[0.5em]
    $\mathbf{B} \perp \mathbf{E}$, $E_0 < B_0$ & $\big(E_0 / B_0,\, \pm \sqrt{1 - (E_0 / B_0)^2}\, ,0\,\big)$ \\
    \addlinespace[0.5em]
    $\mathbf{B} \perp \mathbf{E}$, $E_0 > B_0$ & $\big(B_0 / E_0,\, 0,\, \pm \sqrt{1 - (B_0 / E_0)^2}\,\big)$ \\
    \addlinespace[0.5em]
    $E_0 = 0$ & None; circular orbits on the velocity sphere. \\
    \bottomrule
  \end{tabular}
  \caption{Summary of fixed points $\mathbf{v}^{\pm}$ in velocity space for different field configurations. Here, $E_0$ and $B_0$ are the magnitudes of the electric and magnetic fields, respectively, and $\alpha$ is the relative angle between them.}
  \label{tab:fixed_points}
\end{table}

For positive charges, the fixed points $\mathbf{v}^+$ and $\mathbf{v}^-$ act as an attractor and a repeller, respectively; for negative charges, these roles are reversed.

A similar analysis in configuration space is achieved if we define the compactified coordinates $\xi \equiv t/\mathcal{E}$, $\boldsymbol{\xi}\equiv \mathbf{x}/\mathcal{E}$. A straightforward calculation shows that for $\lambda\rightarrow\pm\infty$ the new variables saturate to
\begin{equation}
\xi^{\pm}=\frac{1}{\boldsymbol{\omega}_E\cdot\mathbf{v}^{\pm}}\,,\qquad\boldsymbol{\xi}^{\pm}=\frac{\mathbf{v}^{\pm}}{\boldsymbol{\omega}_E\cdot\mathbf{v}^{\pm}}\,.
\end{equation}

The speed of the particle asymptotically approaches the speed of light, independently of its mass or initial conditions, so that the long-term dynamics are determined solely by the parameters of the electromagnetic field.

\subsubsection{Numerical results}

The dynamics of the system are sensitive to the ratio of the electric and magnetic field magnitudes \cite{moralesBehaviourChargedSpinning2018}. We analyze numerical results for the magnetically dominant regime, $B_0>E_0$, where intricate orbits arise exhibiting rich dynamics. To start with, we normalize the constants $q=1$ and $\omega_{B}=1$; consequently, the only free parameters are $m \geq 0$, $\omega_{E}\in[0,1]$ and $\alpha\in[0,\pi]$.

\begin{figure*}[t]
  \centering
  \begin{minipage}{0.49\textwidth}
    \centering
    \textbf{(a)}\includegraphics{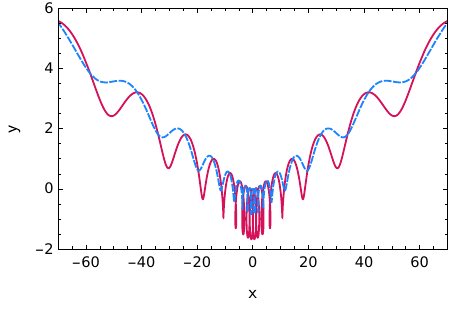}
  \end{minipage}\hfill
  \begin{minipage}{0.49\textwidth}
    \centering
    \textbf{(c)}\includegraphics{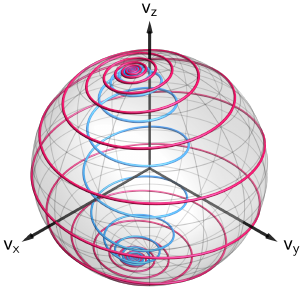}
  \end{minipage}

  \vspace{1em} 

  \begin{minipage}{0.49\textwidth}
    \centering
    \textbf{(b)}\includegraphics[scale=0.95]{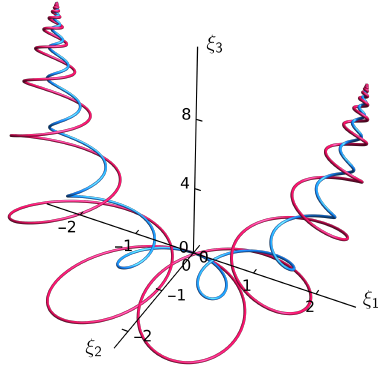}
  \end{minipage}\hfill
  \begin{minipage}{0.49\textwidth}
    \centering
    \textbf{(d)}\includegraphics{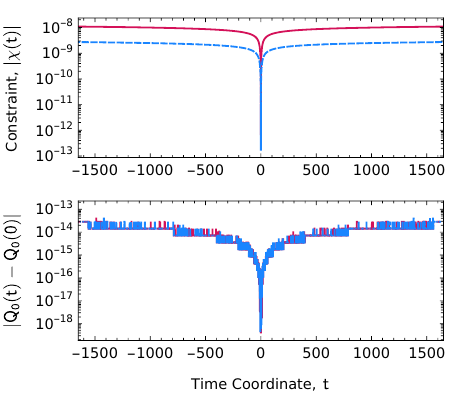}
  \end{minipage}
\caption{
Dynamics of massless and massive charged particles in a constant electromagnetic field.
Parameters: particle charge $q=1$; electric field $\omega_E=0.212$; magnetic field $\omega_B=1$; relative angle between the fields $\alpha = 3\pi/8$.
Mass values: $m=0$ (red), $m=0.8$ (blue).
Initial conditions: energy $\mathcal{E}_0=1$, velocity angles $(\varphi_0, \psi_0) = (\pi/2, 0)$, position $\mathbf{x}_0 = \mathbf{0}$.
(a) Projection of spatial trajectories onto the $z=0$ plane.
(b) Particle trajectories on the velocity sphere; the massive particle explores the interior, whereas the massless one remains confined to the surface. The asymptotic behavior is governed by the fixed points, where the orbits spiral inward or outward.
(c) Spatial trajectories represented in the compactified $\xi$--coordinates.
(d) Evolution of the mass-shell constraint $\chi(t)$ and of the conserved quantity $|Q_0(t) - Q_0(0)|$ as functions of the time coordinate $t$ (the remaining integrals of motion display similar behavior). Smaller deviations from zero indicate better numerical accuracy.
}
\label{fig:massless_vs_massive}
\end{figure*}

\paragraph{Dynamics of Charged Particles: Massive versus Massless}\label{sec:Dynamics of Charged Particles: Massive versus Massless}

For the particular choice of initial conditions, close to the origin of the spatial coordinates, $\mathcal{E}$ remains approximately constant, and the dominant effect of the magnetic field is noticeable, leading to helical trajectories 
(see Fig.~\ref{fig:massless_vs_massive}(a)). Gradually, the orbits are unfolded in the $\hat{\mathbf{x}}$ direction due to the $\mathbf{E} \times \mathbf{B}$ drift force. This is accompanied by an elongation effect caused by the electric field.

For large values of $\lambda$, the electric field becomes predominant, leading to a distinctive elongation of the orbits. Oscillations near the origin are similar for both massive and massless particles (blue dashed and red curves, respectively, in Fig.~\ref{fig:massless_vs_massive}(a)); however, the helical trajectory is more pronounced in the massive case. This occurs because the electric field, while performing work on both types of particles, cannot change the speed of massless ones, thus affecting only their direction, whereas it can still increase the speed of massive particles until it reaches its asymptotic saturation value.

The representation of the orbit in terms of the $\xi$--coordinates (Fig.~\ref{fig:massless_vs_massive}(b)) highlights the universal asymptotic behavior of charged particles, independent of their mass.

We now examine the dynamics in velocity space. As expected, the orbit of a massless particle remains strictly confined to the surface of the velocity sphere, whereas a massive particle can explore its interior (Fig.~\ref{fig:massless_vs_massive}(c)). Qualitatively, the number of cycles described by the orbits is the same in both cases. For arbitrarily large positive (negative) values of $\lambda$, we have $\mathcal{E}\rightarrow\infty$, and both trajectories asymptotically approach the fixed point in the future (past), which can be computed by substituting the specific electromagnetic field parameters into Eqs.~(\ref{eq:vel-space-FixPo}), yielding $\mathbf{v}^{\pm}=\left(0.1945,\pm0.0161,\pm0.9808\right)$.

Thus, we observe that the behavior of massive charged particles approaches that of massless ones as $\mathcal{E}$ increases. The same conclusion holds if the mass is kept fixed while the initial energy $\mathcal{E}_0$ is raised. Reaching this result does not require taking the aforementioned discontinuous limit $m \rightarrow 0$.

\paragraph{Numerical Comparison Between DTF and UTF}

The numerical results of Sec.~\ref{sec:Dynamics of Charged Particles: Massive versus Massless} were obtained using a fourth-order Runge-Kutta (RK4) integrator with a fixed step size $\Delta\lambda = 10^{-3}$ applied to the DTF. The numerical evolution preserves both the mass-shell constraint and the conserved quantities within controlled accuracy (see Fig.~\ref{fig:massless_vs_massive}(d)).

In the DTF, the temporal coordinate evolves nontrivially whenever an electric field is present. For a fixed $\Delta\lambda$, the effective time increment is $\Delta t(\lambda) = \Delta\lambda\mathcal{E}(\lambda)/\mathcal{E}_0$, becoming finer when $\mathcal{E} < \mathcal{E}_0$ and coarser when $\mathcal{E} > \mathcal{E}_0$. This observation prompts a detailed performance comparison between DTF and UTF.

Accordingly, we perform RK4 integrations with $\Delta\lambda = 10^{-3}$ for both frameworks, using the same parameters and initial conditions as in Sec.~\ref{sec:Dynamics of Charged Particles: Massive versus Massless} (see the caption of Fig.~\ref{fig:massless_vs_massive}). For brevity's sake, the analysis is restricted to the massless case, since the results for massive particles are qualitatively similar.

\begin{figure*}[t]
  \centering
  \begin{minipage}{0.49\textwidth}
    \centering
    \textbf{(a)}\includegraphics{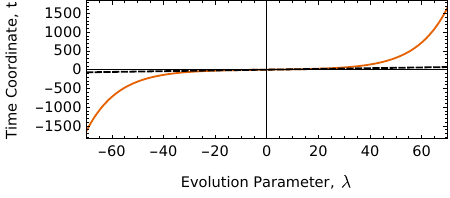}
  \end{minipage}\hfill
  \begin{minipage}{0.49\textwidth}
    \centering
    \textbf{(b)}\includegraphics{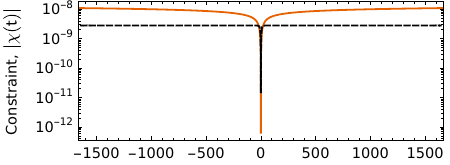}
  \end{minipage}

  \vspace{1em} 

  \begin{minipage}{0.49\textwidth}
    \centering
    \textbf{(c)}\includegraphics{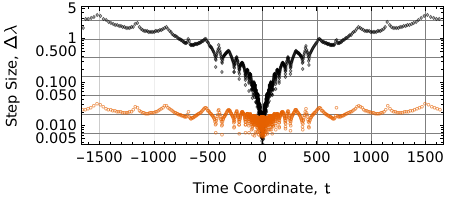}
  \end{minipage}\hfill
  \begin{minipage}{0.49\textwidth}
    \centering
    \textbf{(d)}\includegraphics{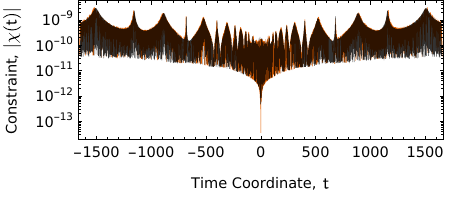}
  \end{minipage}
\caption{
Numerical comparison between the UTF (black) and the DTF (orange) for the massless particle in a constant electromagnetic field.
The parameters and initial conditions are the same as in Fig.~\ref{fig:massless_vs_massive}.
Integration was performed using the RK4 method with $\lambda \in [0,1000]$ for the UTF and $\lambda \in [0,70]$ for the DTF, in both cases with $\Delta\lambda = 10^{-3}$ (all plots were mirrored for negative $\lambda$ values to highlight the symmetry of the solutions).
(a) Evolution of the temporal coordinate $t$ as a function of $\lambda$.
Although $t$ behaves similarly in both frameworks for small $\lambda$, the DTF exhibits exponential growth at large $\lambda$.
(b) Evolution of the mass-shell constraint $\chi(t)$ versus $t$ using the RK4 integrator. (c) Adaptive step size of the RKDP integrator as a function of $t$, comparing the UTF (black $\diamond$) and DTF (orange $\circ$). The UTF step size varies over several orders of magnitude, whereas the DTF step size exhibits significantly smaller fluctuations, remaining around $0.01$.
(d) Evolution of $\chi(t)$ versus $t$ using the adaptive RKDP integrator; in this case, the slight advantage of the UTF over the DTF becomes negligible. 
}
\label{fig:UTF_vs_DTF}
\end{figure*}

A first distinctive feature of the DTF is that the system reaches large values of the effective time $t$ for relatively small values of $\lambda$. As shown in Fig.~\ref{fig:UTF_vs_DTF}(a), $t$ grows linearly for small $|\lambda|$, before entering a regime of exponential growth that far exceeds the uniform-time evolution. More specifically, for $\lambda = 70$, the UTF evolution reaches $t = 70$, while in the DTF it extends up to $t = 1650$.

A second characteristic of the DTF is the adaptive behavior of $\Delta t$. As shown in Fig.~\ref{fig:UTF_vs_DTF}(a), within the interval $-20 \lesssim \lambda \lesssim 20$, the evolution of $t$ slows down, effectively refining the integration precisely where the dynamics are most intricate (corresponding to the helical segment of the orbit in Fig.~\ref{fig:massless_vs_massive}(a)). As the orbit unfolds, $\Delta t$ increases, thereby reducing the computational cost.

A comparison of the mass-shell constraint evolution (Fig.~\ref{fig:UTF_vs_DTF}(b)) reveals good long-term stability for both schemes, with a slight advantage for uniform-time scheme. Achieving this level of accuracy, however, required approximately $10^6$ integration steps for UTF, compared to only $10^4$ for DTF. A fair comparison must therefore take this computational disparity into account. To that end, we repeated the simulations using step sizes of $\Delta\lambda_\mathrm{UTF} = 10^{-2}$ and $\Delta\lambda_\mathrm{DTF} = \Delta\lambda_\mathrm{UTF} \times (70/1650)$, which yielded an equal number of iterations ($\sim10^5$) for both. Under these conditions, DTF proved to be superior by approximately three orders of magnitude. These results are summarized in Table~\ref{tab:DTF_UTF_steps}, which also includes additional tests for several step sizes.
\begin{table}[t]
  \centering
  \setlength{\tabcolsep}{7.5pt} 
  \begin{tabular}{@{}lcccc@{}}
    \toprule
    \makecell{\textbf{Framework}} &
    \makecell{\textbf{Step}\\\textbf{size $\Delta\lambda$}} &
    \makecell{\textbf{Iterations}\\($\lambda_\text{max}/\Delta\lambda$)} &
    \makecell{\textbf{Mass-shell}\\\textbf{violation ($\sim$)}} \\
    \midrule
    DTF & $1\times10^{-2}$ & $7\times10^{3}$ & $10^{-6}$ \\
    \addlinespace[0.5em]
    DTF & $1\times10^{-3}$ & $7\times10^{4}$ & $10^{-8}$ \\
    \addlinespace[0.5em]
    UTF & $1\times10^{-2}$ & $1.7\times10^{5}$ & $10^{-6}$ \\
    \addlinespace[0.5em]
    DTF & $4\times10^{-4}$ & $1.7\times10^{5}$ & $10^{-9}$ \\
    \addlinespace[0.5em]
    UTF & $1\times10^{-3}$ & $1.7\times10^{6}$ & $10^{-9}$ \\
    \addlinespace[0.5em]
    DTF & $4\times10^{-5}$ & $1.7\times10^{6}$ & $10^{-12}$ \\
    \bottomrule
  \end{tabular}
\caption{Performance comparison of RK4 integrations in the dynamical-time (DTF) and uniform-time (UTF) frameworks, showing the step size ($\Delta\lambda$), number of iterations ($\lambda_\text{max}/\Delta\lambda$), and resulting mass-shell violation.}
  \label{tab:DTF_UTF_steps}
\end{table}

Further insight was obtained using a Dormand–Prince (RKDP) integrator with adaptive step size. In this configuration, both frameworks exhibited substantial improvement, and the previous advantage of the uniform-time case was largely reduced (compare the constraint violation in Fig.~\ref{fig:massless_vs_massive}(d) with that in Fig.~\ref{fig:UTF_vs_DTF}(d)). A closer examination of the step-size adaptation, shown in Fig.~\ref{fig:UTF_vs_DTF}(c), reveals that both time parametrizations trace correlated dynamical structures, as evidenced by the pattern of the peaks. However, UTF requires step-size variations spanning several orders of magnitude, with the smallest steps ($\sim 10^{-3}$) occurring in regions where the helical motion is most pronounced. In contrast, DTF maintains an almost uniform step size, with a representative value of order $\sim 10^{-2}$. For the present case study, this property of stable step-size control, combined with the extended effective-time evolution, likely accounts for the superior numerical accuracy of DTF relative to UTF (see Table~\ref{tab:DTF_UTF_steps}).

\begin{figure*}[t]
  \centering
  \begin{minipage}{0.33\textwidth}
    \centering
    \textbf{(a)}\includegraphics[scale=0.87]{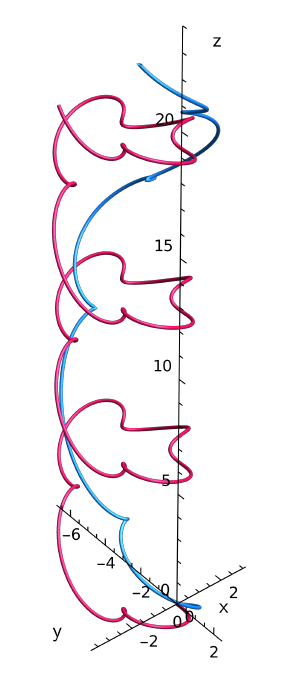}
  \end{minipage}\hfill
  \begin{minipage}{0.33\textwidth}
    \centering
    \textbf{(b)}\includegraphics[scale=0.85]{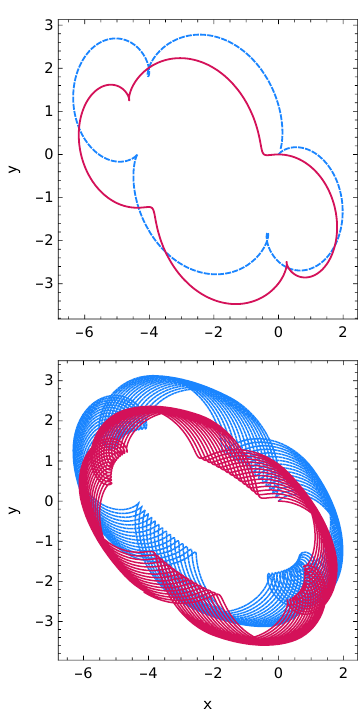}
  \end{minipage}\hfill
  \begin{minipage}{0.33\textwidth}
    \centering
    \textbf{(c)}\includegraphics[scale=0.9]{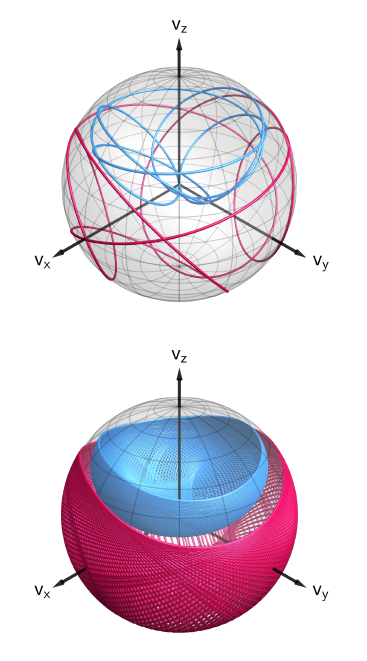}
  \end{minipage}
\caption{
Dynamics of massless and massive charged particles in an elliptically polarized electromagnetic wave with a uniform axial magnetic field.
Mass values: $m=0$ (red), $m=1$ (blue).
Parameters: $q=1$, $a_1=1$, $a_2=1.6180$, $\omega=0.3236$, $\delta=\pi/4$, $b_0=1.6180$.
Initial conditions: massless particle, $\mathbf{p}_0 = (1,0,0)$; massive particle, $\mathbf{p}_0 = \mathbf{0}$; for both cases, $\mathcal{E}_0 = 1$ and $\mathbf{x}_0 = \mathbf{0}$.
(a) Spatial trajectories.
(b) Projection of spatial trajectories onto the $z=0$ plane, showing periodic orbits for $b_0 = 1.6180$ (upper panel) and quasiperiodic ones for $b'_0 = 1.6280$ (lower panel).
(c) Particle trajectories on the velocity sphere; the massive particle explores the interior, whereas the massless one remains confined to the surface; periodic orbits for $b_0 = 1.6180$ (upper panel) and quasiperiodic ones for $b'_0 = 1.6280$ (lower panel).
}
\label{fig:Massive_Massles_EMW}
\end{figure*}

\subsection{Time-dependent electromagnetic field\label{subsec:Campo-oscilante}}
\subsubsection{Numerical results}

As a second case study, we consider an external electromagnetic field configuration that depends explicitly on the time coordinate. Specifically,
\begin{align*}
\mathbf{E}(t,\mathbf{x}) & =a_{1}\cos\left(\omega (t-z)\right)\hat{\mathbf{x}}+a_{2}\cos\left(\omega (t-z)+\delta\right)\hat{\mathbf{y}},
\end{align*}
and the corresponding magnetic field follows from \mbox{$\mathbf{B} = \hat{\mathbf{z}}\times\mathbf{E}$}. These fields describe a monochromatic, elliptically polarized wave propagating along the $\hat{\mathbf{z}}$ direction, where $\omega$ is the frequency, $a_1$ and $a_2$ are the semiaxes of the polarization ellipse, and $\delta$ is the relative phase. Circular polarization corresponds to $a_{1}=a_{2}$ and $\delta=\pi/2$, while linear polarization is obtained for $\delta=\pi$.

A uniform magnetic field $\mathbf{b}=b_{0}\hat{\mathbf{z}}$, parallel to the propagation direction, is also included to increase the dynamical complexity of the system. The particle trajectories in configuration space are displayed in Fig.~\ref{fig:Massive_Massles_EMW}(a). The massive particle exhibits a more pronounced elongation, which can be attributed to its inertia. Although seemingly counterintuitive, it is noteworthy that the massive particle advances ahead of the massless one.

The projection of the trajectories onto the plane perpendicular to the propagation axis shows orbits with similar qualitative features. For $b_0 = 1.6180$, the projected trajectories are periodic (Fig.~\ref{fig:Massive_Massles_EMW}(b), upper panel), whereas small variations in this parameter lead to a quasiperiodic regime (Fig.~\ref{fig:Massive_Massles_EMW}(b), lower panel).

A similar behavior is observed on the velocity sphere: the orbits are periodic for $b_0 = 1.6180$ and become quasiperiodic under small perturbations of this parameter (Fig.~\ref{fig:Massive_Massles_EMW}(c)). Moreover, this representation clearly reveals that the velocity component $v_z$ takes negative values only for the massless particle, while it remains strictly positive for the massive one. Although the massless particle reaches a maximum $|v_z|\sim 1$ during its backward motion, its peak forward velocity is smaller than that of the massive particle. This explains why the massive particle can move ahead: the massless one intermittently retrogrades along the $z$-direction, while the massive particle consistently moves forward.

\begin{figure}[t]
  \begin{minipage}{0.48\textwidth}
  \centering
  \textbf{(a)}\includegraphics{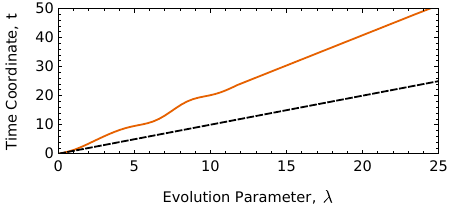}
  \end{minipage}
  \vspace{1em} 
  \begin{minipage}{0.48\textwidth}
  \centering
  \textbf{(b)}\includegraphics{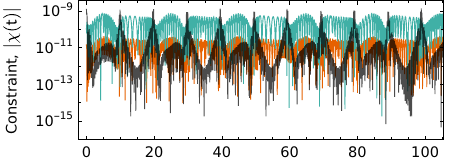}
  \end{minipage}
  \vspace{1em} 
  \begin{minipage}{0.48\textwidth}
  \centering
  \textbf{(c)}\includegraphics{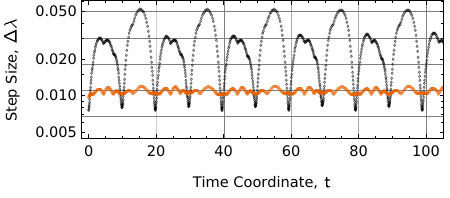}
  \end{minipage}
\caption{
Numerical comparison between the UTF (black) and the DTF (orange or green) for the massive particle in an elliptically polarized electromagnetic wave with a uniform axial magnetic field. The parameters and initial conditions are the same as in Fig.~\ref{fig:Massive_Massles_EMW}.
(a) Evolution of the temporal coordinate $t$ as a function of the parameter $\lambda$.
(b) Evolution of the mass-shell constraint $\chi(t)$ versus $t$, comparing UTF with $\Delta\lambda=10^{-2}$ (black), DTF with $\Delta\lambda=10^{-2}$ (green), and DTF with $\Delta\lambda=5\times10^{-3}$ (orange); smaller deviations from zero indicate better numerical accuracy.
(c) Adaptive step size of the RKDP integrator as a function of $t$, comparing the UTF (black $\diamond$) and DTF (orange $\circ$). In both cases, the step size oscillates with similar behavior, but the amplitude is markedly smaller for DTF.
}
\label{fig:UDT_vs_DTF}
\end{figure}
As shown in Fig.~\ref{fig:UDT_vs_DTF}(a), the dynamical time advances roughly twice as fast as the uniform time (a milder effect than in the previous case study). This occurs because the oscillating electric field generates alternating accelerations and decelerations in the progression of the time coordinate. The discrepancies in time progression are significant when evaluating the numerical performance of the two systems, as discussed below.

To this end, we monitored the violation of the mass-shell condition.  Figure \ref{fig:UDT_vs_DTF}(b) presents the outcomes derived from the RK4 integrator utilizing a constant step size of $\Delta\lambda = 10^{-2}$, with the black and green lines representing UTF and DTF, respectively. Both schemes exhibit oscillatory behavior; however, the UTF curve has pronounced peaks, while the DTF profile is somewhat smoother. Due to the faster progression of DTF, which occurs at about double the rate, UTF necessitates roughly double the number of iterations to cover the same physical time. To compensate for this disparity, we reduced the DTF step size accordingly. The resulting curve (orange line in Fig.~\ref{fig:UDT_vs_DTF}(b)) shows the best overall performance.

We also applied the adaptive RKDP integrator, as in the previous case study, to further understand the role of the dynamical time in the simulations. In this case, the integration step size oscillates with a similar frequency for both schemes (Fig.~\ref{fig:UDT_vs_DTF}(c)); however, the amplitude is considerably smaller and more regular in DTF. This behavior reinforces the previous finding that the dynamical-time parametrization provides a more adaptive framework for capturing the system’s underlying dynamics.

\section{Concluding Remarks}\label{sec:Conclusions}

We have proposed a dynamical framework to describe the motion of charged test particles, both massive and massless, in external electromagnetic fields. Our formulation relies on two key ingredients: a specific gauge choice for the einbein, which promotes time and energy to dynamical variables, and the \emph{free} evolution of an eight-dimensional dynamical system, $(t,\mathbf{x};\mathcal{E},\mathbf{P})$, subject to initial data satisfying the mass-shell constraint. The variables introduced here should not be confused with the canonical variables of the Hamiltonian formalism. Not all degrees of freedom are physical or independent, since reparametrization invariance renders the evolution of $t$ physically irrelevant, and the mass-shell constraint requires only three of the four variables $(\mathcal{E},\mathbf{P})$ to define a well-posed initial-value problem. Consequently, the present framework provides a unified description of massive and massless sectors, distinguished solely by initial conditions encoding the particle's mass.

This should be contrasted with constrained evolution, in which one typically adopts the uniform-time gauge, solves the mass-shell constraint to determine $\mathcal{E} = \sqrt{m^2 + \mathbf{P}^2}$, and then evolves the remaining six variables using the evolution equations. It should be noted that this procedure, in principle, forbids negative-energy solutions, which are allowed classically for massless particles and underlie the so-called Klein tunneling phenomenon in quantum theory \cite{moralesQuantumChargedSpinning2019}.

Promoting time to a dynamical variable renders the system autonomous, while including energy eliminates the need for the Lorentz factor, a quantity that becomes ill-defined in the massless limit and numerically unstable in the ultrarelativistic regime.

From the initial-value formulation of the classical system, it follows that free, spinless, massless particles moving in the same direction and passing through the same spacetime point at $t(0)$ can be distinguished solely by their energy $\mathcal{E}_0$. This effective role of energy as inertia for massless particles, which was noted in Ref.~\cite{moralesBehaviourChargedSpinning2018}, is confirmed by our numerical results. Invoking the correspondence principle, one may suggest that such particles are distinguished by their de Broglie wavelengths, $\lambda_{\mathrm{dB}} = hc/\mathcal{E}_0$.\footnote{A general reformulation including spin degrees of freedom is required to further substantiate this correspondence. We thank Alexei Deriglazov for drawing our attention to this subtle point.}

We have also shown that the velocity-space representation provides an alternative perspective on charged-particle dynamics. The velocity sphere defines a submanifold that constrains the motion of massless particles, while its interior is accessible only to massive ones. This representation allows a clear characterization of asymptotic behaviors. Furthermore, with an angular parametrization of the velocity sphere, initial conditions consistent with the mass-shell constraint can be readily specified.

The evolution of time, being proportional to the particle's energy, arises from a specific parametrization choice, which we term the dynamical-time gauge. A notable feature of this gauge is that the subset of equations governing momentum and energy remains linear, although the coefficients depend locally on the spatial and temporal coordinates. Whether alternative gauge choices could provide analytical or numerical advantages remains an open question for future work.

From a numerical perspective, the dynamical-time framework (DTF) exhibits distinctive advantages over the conventional uniform-time framework (UTF). Both schemes provide accurate simulations, irrespective of the particle's mass, with only minor violations of the mass-shell constraint and conserved quantities. Notably, as suggested by our numerical results, when electric fields directly drive the evolution of the time coordinate, the effective time variable adapts dynamically, reaching stages of evolution far exceeding those achievable in the uniform-time framework. This adaptive clock rate closely resembles the behavior of a variable-step integrator operating within a uniform-time framework. These results, verified for both uniform and time-dependent electromagnetic field configurations, constitute one of the main numerical findings of this work.

Overall, the proposed framework represents a viable and efficient alternative for numerical simulations. Despite the simplifications of our model, its ability to encompass the massless limit provides a consistent extension of charged-particle dynamics into the ultrarelativistic regime, offering new opportunities for both analytical and numerical exploration.

\acknowledgments{
We thank Alexei Deriglazov for extensive and insightful correspondence related to this work. We are also grateful to Olivier Piguet for a careful reading of the manuscript and for several valuable suggestions, and to Lucy Aduviry for bringing Ref.~\cite{LeBellac:1973unm} to our attention. GMRA acknowledges support from the European Union’s Horizon 2020 research and innovation program under the Marie Sklodowska-Curie grant agreement No.~101034383.
}
\bibliography{cas-refs.bib}

@article{akhiezerDynamicsHighenergyCharged1995,
  title = {Dynamics of High-Energy Charged Particles in Straight and Bent Crystals},
  author = {Akhiezer, A I and Shul'ga, N F and Truten', V I and Grinenko, A A and Syshchenko, V V},
  year = {1995},
  month = oct,
  journal = {Phys.-Usp.},
  volume = {38},
  number = {10},
  pages = {1119--1145},
  issn = {1063-7869, 1468-4780},
  doi = {10.1070/PU1995v038n10ABEH000114},
  urldate = {2025-03-27}
}

@article{balachandranClassicalDescriptionParticle1977,
  title = {Classical Description of a Particle Interacting with a Non-{{Abelian}} Gauge Field},
  author = {Balachandran, A. P. and Salomonson, Per and Skagerstam, Bo Sture and Winnberg, Jan Olov},
  year = {1977},
  journal = {Phys. Rev. D},
  volume = {15},
  number = {8},
  pages = {2308--2317},
  issn = {05562821},
  doi = {10.1103/PhysRevD.15.2308}
}

@article{botetDualityNonHermitianTwostate2019,
  title = {The Duality between a Non-{{Hermitian}} Two-State Quantum System and a Massless Charged Particle},
  author = {Botet, Robert and Kuratsuji, Hiroshi},
  year = {2019},
  month = dec,
  journal = {J. Phys. A: Math. Theor.},
  volume = {52},
  number = {3},
  pages = {35303},
  issn = {17518121},
  doi = {10.1088/1751-8121/aaf479}
}

@article{brinkLagrangianFormulationClassical1977,
  title = {A {{Lagrangian}} Formulation of the Classical and Quantum Dynamics of Spinning Particles},
  author = {Brink, L. and Di Vecchia, P. and Howe, P.},
  year = {1977},
  journal = {Nucl. Phys. B},
  volume = {118},
  number = {1-2},
  pages = {76--94},
  issn = {05503213},
  doi = {10.1016/0550-3213(77)90364-9}
}

@article{brinkLocalSupersymmetrySpinning1976,
  title = {Local Supersymmetry for Spinning Particles},
  author = {Brink, L. and Deser, S. and Zumino, B. and Vecchia, P. Di and Howe, P.},
  year = {1976},
  journal = {Phys. Lett. B},
  volume = {64},
  number = {4},
  pages = {435--438},
  issn = {0370-2693},
  doi = {10.1016/0370-2693(76)90115-5}
}

@book{buchnerSpaceAstrophysicalPlasma2023,
  title = {Space and {{Astrophysical Plasma Simulation}}: {{Methods}}, {{Algorithms}}, and {{Applications}}},
  shorttitle = {Space and {{Astrophysical Plasma Simulation}}},
  author = {B{\"u}chner, J{\"o}rg},
  year = {2023},
  edition = {1st ed},
  publisher = {Springer},
  address = {Cham},
  isbn = {978-3-031-11870-8},
  doi = {https://doi.org/10.1007/978-3-031-11870-8},
  langid = {english}
}

@book{buchnerSpacePlasmaSimulation2003,
  title = {Space {{Plasma Simulation}}},
  editor = {B{\"u}chner, J{\"o}rg and Dum, Christian T. and Scholer, Manfred},
  year = {2003},
  series = {Lecture {{Notes}} in {{Physics}}},
  volume = {615},
  publisher = {Springer},
  address = {Berlin},
  doi = {10.1007/3-540-36530-3},
  urldate = {2025-03-28},
  copyright = {http://www.springer.com/tdm},
  isbn = {978-3-540-00698-5 978-3-540-36530-3},
  langid = {english}
}

@article{castronetoElectronicPropertiesGraphene2009,
  title = {The Electronic Properties of Graphene},
  author = {Castro Neto, A. H. and Guinea, F. and Peres, N. M.R. and Novoselov, K. S. and Geim, A. K.},
  year = {2009},
  month = jan,
  journal = {Rev. Mod. Phys.},
  volume = {81},
  number = {1},
  pages = {109--162},
  issn = {00346861},
  doi = {10.1103/RevModPhys.81.109}
}

@article{dealcantarabonfimChaoticHyperchaoticMotion2000,
  title = {Chaotic and Hyperchaotic Motion of a Charged Particle in a Magnetic Dipole Field},
  author = {De Alcantara Bonfim, O. F. and Griffiths, David J. and Hinkley, Sasha},
  year = {2000},
  month = jan,
  journal = {Int. J. Bifurcation Chaos},
  volume = {10},
  number = {1},
  pages = {265--271},
  issn = {02181274},
  doi = {10.1142/S0218127400000177}
}

@article{dunneTopologicalChernSimonsQuantum1990,
  title = {``{{Topological}}'' ({{Chern-Simons}}) Quantum Mechanics},
  author = {Dunne, G. V. and Jackiw, R. and Trugenberger, C. A.},
  year = {1990},
  month = jan,
  journal = {Phys. Rev. D},
  volume = {41},
  number = {2},
  pages = {661--666},
  doi = {10.1103/PhysRevD.41.661}
}

@article{friedmanRelativisticAccelerationCharged2005,
  title = {Relativistic Acceleration of Charged Particles in Uniform and Mutually Perpendicular Electric and Magnetic Fields as Viewed in the Laboratory Frame},
  author = {Friedman, Yaakov and Semon, Mark D.},
  year = {2005},
  month = aug,
  journal = {Phys. Rev. E},
  volume = {72},
  number = {2},
  pages = {026603},
  doi = {10.1103/PhysRevE.72.026603}
}

@article{geimGrapheneExploringCarbon2007,
  title = {Graphene: {{Exploring}} Carbon Flatland},
  author = {Geim, Andrey K. and MacDonald, Allan H.},
  year = {2007},
  month = aug,
  journal = {Phys. Today},
  volume = {60},
  number = {8},
  pages = {35--41},
  issn = {00319228},
  doi = {10.1063/1.2774096}
}

@book{henneauxQuantizationGaugeSystems1992,
  title = {Quantization of {{Gauge Systems}}},
  author = {Henneaux, Marc and Teitelboim, Claudio},
  year = {1992},
  eprinttype = {jstor},
  publisher = {Princeton University Press},
  urldate = {2024-09-29},
  isbn = {978-0-691-08775-7}
}

@article{hosurRecentDevelopmentsTransport2013,
  title = {Recent Developments in Transport Phenomena in {{Weyl}} Semimetals},
  author = {Hosur, Pavan and Qi, Xiaoliang},
  year = {2013},
  month = nov,
  journal = {C. R. Phys.},
  volume = {14},
  number = {9-10},
  pages = {857--870},
  issn = {16310705},
  doi = {10.1016/j.crhy.2013.10.010}
}

@book{kieferQuantumGravity2004,
  title = {Quantum {{Gravity}}},
  author = {Kiefer, Claus},
  year = {2004},
  publisher = {Oxford University Press UK}
}

@article{moralesBehaviourChargedSpinning2018,
  title = {Behaviour of {{Charged Spinning Massless Particles}}},
  author = {Morales, Ivan and Neves, Bruno and Oporto, Zui and Piguet, Olivier},
  year = "2018",
  journal = "Symmetry",
  volume = "10",
  number = "1",
  issn = "2073-8994",
  url = "https://doi.org/10.3390/sym10010002",
}

@article{moralesQuantumChargedSpinning2019,
  title = {Quantum Charged Spinning Massless Particles in 2 + 1 Dimensions},
  author = {Morales, Ivan and Neves, Bruno and Oporto, Zui and Piguet, Olivier},
  year = {2019},
  month = dec,
  journal = {Eur. Phys. J. C},
  volume = {79},
  number = {12},
  issn = {14346052},
  url = { https://doi.org/10.1140/epjc/s10052-019-7511-z},
}

@article{petriFullyImplicitNumerical2017,
  title = {A Fully Implicit Numerical Integration of the Relativistic Particle Equation of Motion},
  author = {P{\'e}tri, J.},
  year = {2017},
  month = apr,
  journal = {J. Plasma Phys.},
  volume = {83},
  number = {2},
  pages = {705830206},
  issn = {0022-3778},
  doi = {10.1017/S0022377817000307}
}

@article{ramDynamicsChargedParticles2010,
  title = {Dynamics of Charged Particles in Spatially Chaotic Magnetic Fields},
  author = {Ram, Abhay K. and Dasgupta, Brahmananda},
  year = {2010},
  month = dec,
  journal = {Phys. Plasma},
  volume = {17},
  number = {12},
  pages = {122104},
  issn = {1070-664X, 1089-7674},
  doi = {10.1063/1.3529366},
  urldate = {2025-03-27},
  langid = {english}
}

@article{ruggieroTaleAnalogiesReview2023,
  title = {A Tale of Analogies: A Review on Gravitomagnetic Effects, Rotating Sources, Observers and All That},
  shorttitle = {A Tale of Analogies},
  author = {Ruggiero, Matteo Luca and Astesiano, Davide},
  year = {2023},
  month = nov,
  journal = {J. Phys. Commun.},
  volume = {7},
  number = {11},
  pages = {112001},
  issn = {2399-6528},
  doi = {10.1088/2399-6528/ad08cf},
  urldate = {2025-03-28}
}

@article{saavedraDegenerateDynamicalSystems2001,
  title = {Degenerate Dynamical Systems},
  author = {Saavedra, Joel and Troncoso, Ricardo and Zanelli, Jorge},
  year = {2001},
  month = sep,
  journal = {J. Math. Phys.},
  volume = {42},
  number = {9},
  pages = {4383--4390},
  issn = {0022-2488},
  doi = {10.1063/1.1389088}
}

@article{shebalinStormerRegionsAxisymmetric2004,
  title = {St{\o}rmer Regions for Axisymmetric Magnetic Multipole Fields},
  author = {Shebalin, John V.},
  year = {2004},
  journal = {Phys. Plasma},
  volume = {11},
  number = {7},
  pages = {3472--3482},
  issn = {1070664X},
  doi = {10.1063/1.1752931}
}

@article{takeuchiSatoshi2002,
  title = {Relativistic {$\mathbf{E} \times \mathbf{B}$} acceleration},
  author = {Takeuchi, Satoshi},
  year = {2002},
  month = sep,
  journal = {Phys. Rev. E},
  volume = {66},
  number = {3},
  pages = {037402},
  doi = {10.1103/PhysRevE.66.037402}
}

@article{xiePeriodQuasiperiodChaos2020,
  title = {From Period to Quasiperiod to Chaos: {{A}} Continuous Spectrum of Orbits of Charged Particles Trapped in a Dipole Magnetic Field},
  author = {Xie, Yuxin and Liu, Siming},
  year = {2020},
  journal = {Chaos},
  volume = {30},
  number = {12},
  issn = {10897682},
  url = {https://doi.org/10.1063/5.0028644},
  pmid = {33380059}
}

@article{eli_1974,
    title={Motion of charged particles in homogeneous electromagnetic fields},
    author={Honig, Eli and Schucking, Engelbert L. and Vishveshwara, C. V.},
    volume={15},
    url={http://dx.doi.org/10.1063/1.1666728},
    doi={10.1063/1.1666728},
    number={6},
    journal={J. Math. Phys.},
    publisher={AIP Publishing},
    year={1974},
    month={Jun},
    pages={774-781}
}

@article{ProccIWOSP,
    author = {Oporto, Zui and Ramírez-Ávila, Gonzalo Marcelo},
    title = {Dynamical analysis of massless charged particles},
    journal = {AIP Conf. Proc.},
    volume = {2731},
    number = {1},
    pages = {040006},
    year = {2023},
    month = {05},
    issn = {0094-243X},
    doi = {10.1063/5.0133494},
    url = {https://doi.org/10.1063/5.0133494}
}

@article{LeBellac:1973unm,
    author = "Le Bellac, M. and L{\'e}vy-Leblond, J. M.",
    title = "{Galilean electromagnetism}",
    doi = "10.1007/BF02895715",
    journal = "Nuovo Cim. B",
    volume = "14",
    number = "2",
    pages = "217--234",
    year = "1973"
}

@article{arreaga-garcia2014equations,
  author = {Guillermo Arreaga-García and Julio Saucedo-Morales},
  title = {Equations of motion of a relativistic charged particle with curvature dependent actions},
  journal = {Palestine J. Math},
  volume = {3},
  number = {2},
  pages = {218--230},
  year = {2014},
  note = {Also available as arXiv:1308.4714 [physics.class-ph]},
  url = {https://pjm.ppu.edu/paper/82}
}

@article{Plyushchay:1988ws,
    author = "Plyushchay, M. S.",
    title = "{Massive Relativistic Point Particle With Rigidity}",
    reportNumber = "IFVE-88-172",
    doi = "10.1142/S0217751X89001564",
    journal = "Int. J. Mod. Phys. A",
    volume = "4",
    pages = "3851",
    year = "1989"
}

@article{Plyushchay:1988wx,
    author = "Plyushchay, M. S.",
    title = "{Massless Point Particle With Rigidity}",
    reportNumber = "IFVE-88-160 (In Russian)",
    doi = "10.1142/S0217732389000988",
    journal = "Mod. Phys. Lett. A",
    volume = "4",
    pages = "837--847",
    year = "1989"
}

@article{Nesterenko:1991av,
    author = "Nesterenko, V. V.",
    title = "{Curvature and torsion of the world curve in the action of the relativistic particle}",
    doi = "10.1063/1.529494",
    journal = "J. Math. Phys.",
    volume = "32",
    pages = "3315--3320",
    year = "1991"
}

@book{Deriglazov2017,
  author = {Deriglazov, Alexei},
  title = {Classical mechanics: Hamiltonian and Lagrangian formalism},
  publisher = {Springer},
  year = {2017},
  edition = {2nd ed.},
  isbn = {978-3-319-44146-7}
}
\newpage{}
\end{document}